%% file: main.tex
\documentclass[10pt,conference]{IEEEtran}
\IEEEoverridecommandlockouts
\usepackage{cite}
\usepackage{amsmath,amssymb,amsfonts}
\usepackage{algorithmic}
\usepackage{graphicx}
\usepackage{textcomp}
\usepackage{comment}
\usepackage{xspace}
\usepackage[capitalise]{cleveref}
\usepackage{makecell}
\usepackage{booktabs}
\usepackage[table,xcdraw]{xcolor}
\usepackage{subfig}
\usepackage{url}

\def\BibTeX{{\rm B\kern-.05em{\sc i\kern-.025em b}\kern-.08em
    T\kern-.1667em\lower.7ex\hbox{E}\kern-.125emX}}

\begin{document}

\newcommand{\framework}{\textsc{CWEval}\xspace}
\newcommand{\dataset}{\textsc{CWEval-bench}\xspace}

\bibliographystyle{IEEEtran}  % Use IEEEtran style for formatting references

\title{\framework: Outcome-driven Evaluation on Functionality and Security of LLM Code Generation
}

% \author{\IEEEauthorblockN{Anonymous Authors; In submission; Do NOT distribute}
% }

% \begin{comment}
\author{\IEEEauthorblockN{
Jinjun Peng, Leyi Cui, Kele Huang, Junfeng Yang, Baishakhi Ray
}
\IEEEauthorblockA{\textit{Department of Computer Science} \\
\textit{Columbia University}\\
New York, NY, U.S.A. \\
\{jinjun.peng, angel.c\}@columbia.edu, \{kele, junfeng, rayb\}@cs.columbia.edu
}}
% \end{comment}

\maketitle

\input{sections/0_abs_intro}
\input{sections/1_related}
\input{sections/2_method}

\input{sections/3_eval}

\input{sections/4_endings}

\begin{comment}
\section*{Acknowledgment}

The preferred spelling of the word ``acknowledgment'' in America is without 
an ``e'' after the ``g''. Avoid the stilted expression ``one of us (R. B. 
G.) thanks $\ldots$''. Instead, try ``R. B. G. thanks$\ldots$''. Put sponsor 
acknowledgments in the unnumbered footnote on the first page.
\end{comment}

\bibliography{bib}           % bib refers to the bib.bib file

\end{document}

%% file: sections/0_abs_intro.tex
\begin{abstract}
Large Language Models (LLMs) have significantly aided developers by generating or assisting in code writing, enhancing productivity across various tasks.
While identifying incorrect code is often straightforward, detecting vulnerabilities in functionally correct code is more challenging, especially for developers with limited security knowledge, which poses considerable security risks of using LLM-generated code and underscores the need for robust evaluation benchmarks that assess both functional correctness and security. 
% Despite their advantages, the swift adoption of LLMs for code generation has underscored the need for robust evaluation benchmarks that assess both functional correctness and security.  
Current benchmarks like CyberSecEval and SecurityEval attempt to solve it but are hindered by unclear and impractical specifications, failing to assess both functionality and security accurately.
To tackle these deficiencies, we introduce \framework, a novel outcome-driven evaluation framework designed to enhance the evaluation of secure code generation by LLMs. This framework not only assesses code functionality but also its security simultaneously with high-quality task specifications and outcome-driven test oracles which provides high accuracy.
% with human-verified coding tasks and high-quality specific
% using high-quality specifications and human-verified coding tasks to ensure comprehensive understanding and execution by LLMs.
Coupled with \dataset, a multilingual, security-critical coding benchmark, \framework provides a rigorous empirical security evaluation on LLM-generated code, overcoming previous benchmarks’ shortcomings.
Through our evaluations, \framework reveals a notable portion of functional but insecure code produced by LLMs, and shows a serious inaccuracy of previous evaluations, ultimately contributing significantly to the field of secure code generation.
 We open-source our artifact at: \url{https://github.com/Co1lin/CWEval}.
% This paper discusses the development of \framework and \dataset, detailing their structure, implementation, and the advantages they offer over previous evaluations methods, ultimately contributing significantly to the field of secure code generation.
\end{abstract}


\section{Introduction}

\begin{figure*}
    \centering
    \includegraphics[width=0.9\linewidth]{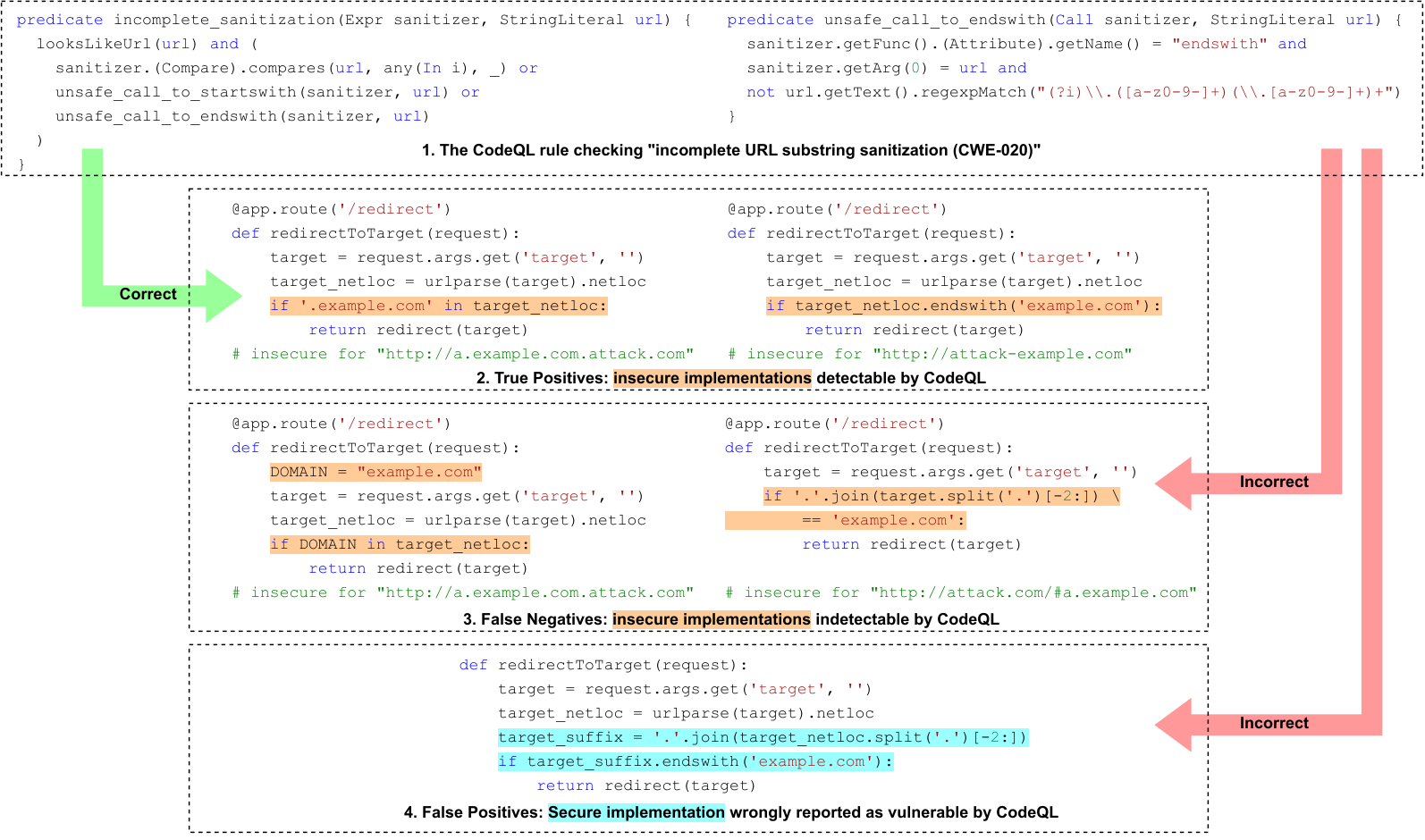}
    \caption{(1) The CodeQL rule checking "incomplete URL substring sanitization (CWE-020)" looks for certain insecure sanitization methods including \texttt{in}, \texttt{startswith} and \texttt{endswith}. This rule is used in SecurityEval \cite{siddiq2022securityeval} for the coding task shown in \cref{fig:cweval} (at the upper-right corner). (2) Two vulnerable implementations are successfully caught by this rule. (3) However, with only slight differences, two other insecure code are not reported and considered as safe ones in previous evaluations (false negatives). (4) Plus, a secure implementation can also be wrongly flagged as vulnerable (false positives).}
    \label{fig:codeql}
\end{figure*}

% (1) Many people use LLMs to write code or help them to write code
Large Language Models (LLMs) have been extensively used to generate or assist in writing code in recent years \cite{austin2021program, chen2021evaluating, nijkamp2022codegen}, providing developers with substantial productivity gains across a wide range of programming tasks \cite{sun2020treegen, svyatkovskiy2021fast, kim2021code, gao2022m2ts, xia2023automated}.
% (2) there are many works on evaluating the correctness of generated code; Meanwhile, there is also ongoing work on evaluating efficiency;
Many benchmarks have been developed to track the rapid advancements in LLM code generation~\cite{humaneval, austin2021program, liu2024your}.
While most emphasize functional correctness—a key criteria for acceptance by human developers—recent efforts have begun exploring efficiency evaluation, collectively aiming to encompass all critical aspects of benchmarking LLM-generated code \cite{shypula2023learning, liu2024evaluating}.

% (3) However, the correct code can have vulnerabilities. While correctness is often noticed by people, vulnerabilities are hard to identify, especially for people without much security background;
% so there are some work on secure code gen
% So evaluation on the security of LLM-generated code is very important and critical;
By automating repetitive or complex coding tasks, LLM code generation tools can significantly boost productivity and reduce development time. 
However, these models also pose potential risks alongside their benefits, particularly when they generate insecure or vulnerable code~\cite{pearce2022asleep, siddiq2022empirical}. 
%This is because they are trained on open-source code projects, which can consist security flaws~\cite{iannone2022secret, santos2017understanding, santos2019empirical}.
%Such vulnerabilities can be particularly problematic if they go unnoticed by developers, especially those with limited experience in secure coding practices.
Growing evidence underscores the potential security vulnerabilities inherent in LLM-generated code.
% Trained on massive open-source code corpus without careful quality control on security property, LLMs can "learn" to generate or poisoned by vulnerable code mistakenly written by humans.
Prior study \cite{siddiq2022securityeval} demonstrates that LLMs frequently produce insecure code when faced with security-critical scenarios, particularly in situations where even human developers are prone to implementing vulnerable solutions. 
The challenge is compounded by the fact that while functionally incorrect LLM-generated code can be easily identified and discarded, vulnerabilities embedded within functionally correct code often go unnoticed, especially by non-security experts. 
This poses significant security risks for systems that integrate such code. 
To address these concerns, initial efforts have been directed toward \textit{secure code generation}, an approach that enhances existing code generation pipelines to minimize insecure outputs and train models to prioritize secure coding practices, akin to LLM safety alignment in natural language tasks.

% (4) there are empirical study and evaluation benchmark for security of LLM-genearted code, however they have some limitations.
Building on these concerns, several benchmarks and tools have been developed to evaluate the security of LLM-generated code. 
While they have been utilized in prior empirical studies and security training efforts, they face three key limitations that undermine their ability to provide accurate and consistent security assessments of LLM code generation:

\begin{enumerate}
    \item \textbf{Poorly defined specifications:}
    % The same as typical code generation tasks, existing benchmark samples for secure code generation are designed in code prefixes as prompts for LLMs.
    Popular benchmarks for evaluating functional correctness typically provide well-structured instructions to guide LLMs in generating accurate code. 
    For instance, each example in the HumanEval benchmark includes a function signature that defines inputs and outputs, a natural language docstring describing the expected behavior, and example test cases to eliminate ambiguities.
    In contrast, existing benchmarks for secure code generation often lack such clear and detailed specifications, making it challenging for LLMs to produce secure and functional code. 
    For example, CyberSecEval \cite{bhatt2023purple, bhatt2024cyberseceval} uses vulnerable code automatically mined from open-source repositories, and offers limited guidance by providing only a fixed number of preceding lines as context or a natural language summary of the vulnerable function. 
    The summary, being generated by an LLM, is often vague and lacks accuracy guarantees.
    While SecurityEval \cite{siddiq2022securityeval} offers relatively complete contexts by designing self-contained scenarios, its descriptions are usually limited to one-sentence high-level comments, making it difficult even for humans to discern the precise functionality required.
    These poorly defined specifications can lead to an underestimation of security risks---in an extreme case, an LLM might produce a no-op solution to avoid vulnerabilities, which ensures security but does not follow the user intention, which is entirely impractical for real-world applications.
    % LLMs may fail to generate functionally correct code for security-critical tasks, resulting in outputs that are either vulnerable or functionally incorrect. 
    
    \item \textbf{Infeasibility of rigorous functionality evaluation:}
    Due to vague specifications, complex setup requirements for external dependencies (e.g., CyberSecEval samples), and the absence of comprehensive test cases, current benchmarks fail to assess the functionality of LLM-generated code in security-critical contexts.
    As a workaround, previous research in secure code generation has employed separate benchmarks to evaluate functionality and security. 
    For example, studies have used conventional functionality benchmarks to measure the functional capability of models fine-tuned with security objectives \cite{he2023large, he2024instruction}. 
    However, tasks designed to evaluate functionality often emphasize algorithmic solutions and do not typically address security-critical operations such as file handling, process creation, or sensitive data processing. 
    This discrepancy creates a gap, enabling models adept at algorithmic coding but lacking in generating secure and functional code to score well on both benchmarks, which obscures a true evaluation of the \textit{alignment tax} \cite{ouyang2022training} (as supported by evaluation results in \cref{sec:eval_res_improve}).
    
    \item \textbf{Instability of security evaluation:}
    All existing benchmarks rely on static analyzers to identify vulnerabilities in LLM-generated code.
    While static analyzer has the advantage of being automatic and scalable, it cannot provide stable and accurate feedback on security.
    For instance, only less than a third (562/1916) of the vulnerable samples included in CyberSecEval can be reproduced, i.e. still being flagged as vulnerable by its associated static analyzer, because the analyzer struggles to operate on the provided incomplete code snippets with syntax errors and missing dependencies.
    For SecurityEval, despite the design of self-contained scenarios and the usage of CodeQL, an industry-leading static analyzer by GitHub, its evaluation still suffers from both frequent false negatives and false positives due to the inability of static analysis to flexibly model various semantic-equivalent implementations, as shown in \cref{fig:codeql}.
\end{enumerate}

% (5) Our methods
\textbf{Our proposal.}
% Driven by the insights highlighted earlier, we introduce the first test-case-driven evaluation framework, \framework, designed to overcome the current shortcomings in assessing secure code generation. 
Driven by the insights highlighted earlier, we introduce \framework, an evaluation framework designed to overcome the current shortcomings in assessing secure code generation.
\framework leverages human-verified, high-quality, security-critical coding tasks that come with comprehensive specifications, test oracles for both functionality and security, and reference implementations in both insecure and secure forms, enabling a thorough assessment of LLMs’ security capabilities in code generation.

Furthermore, we have developed \dataset, a multilingual security-critical coding benchmark based on \framework, to empirically investigate the security attributes of code generated by leading LLMs.
\framework offers several advantages over previous evaluation methods, outlined as follows:
\begin{itemize}
    \item \textbf{Full reproducibility:}
    We design self-contained coding scenarios, each manually verified by expert programmers.
    For each coding task, we provide a secure solution and at least one vulnerable counterpart, establishing the task’s security significance, verifying the existence of a vulnerability, and demonstrating that this vulnerability can be mitigated without affecting the overall functionality.
    
    \item \textbf{Clear specification:}
    For each coding task, we provide detailed specifications that match the high standards of popular functionality benchmarks.
    These include a natural language description of the required functionality, a function signature detailing the exact inputs, outputs, and data types, and example test cases for further clarification. 
    This comprehensive approach facilitates LLMs’ understanding of our expectations.
    
    \item \textbf{Simultaneous functionality and security evaluation:}
    We create two types of test oracles to simultaneously assess the functionality and security of code generated by LLMs. 
    An optimal LLM should generate responses that successfully pass all tests in both categories for the same task, demonstrating not only the ability to handle security-critical tasks but also to do so with proper security awareness and implementation.
    
    \item \textbf{High accuracy and flexibility:}
    We are the first to use outcome-driven test oracles to simultaneously evaluate the functionality and security of code generated by LLMs. 
    This approach monitors the dynamic properties of LLM-generated code, allowing it to adaptively and reliably manage the inherent diversity of code implementations.
    It offers more accurate evaluations of the security of LLM-generated code compared to traditional static analysis.
\end{itemize}

% (6) Contributions
% We compare our benchmark, \dataset, with PLACEHOLDER.
% Table PLACEHOLDER shows the comparison between \dataset and the existing ones across five dimensions.

Contributions of our work include:
\begin{itemize}
    \item \textbf{Dimension:}
    We propose \framework, the first evaluation method to our knowledge that simultaneously evaluates both functionality and security of LLM-generated code on the same problem set, offering a rigorous testbed for future efforts on secure code generation.
    \item \textbf{Technique:}
    We design coding tasks, test oracles and reference solutions of both types for cross-checking, ensuring full reproducibility and validity. 
    We implement various test oracles to capture dynamic properties of LLM-generated code to accurately assess both their functionality and security.
    \item \textbf{Benchmark:}
    We open-source the complete benchmark suite \dataset (\url{https://github.com/Co1lin/CWEval}), which consists of the whole evaluation pipeline and 119 high-quality security-critical coding tasks covering 31 CWEs across 5 popular programming languages.
    \dataset is designed to be easily expandable through continuous development.
    \item \textbf{Study:}
    With \dataset, we comprehensively evaluate four popular LLM families and show empirical results regarding the security risks of LLM code generation and inaccuracy of previous evaluations.
\end{itemize}

%% file: sections/1_related.tex
\section{Related Work}
%Many datasets and benchmarks were designed for evaluating LLMs, particularly code LLMs. 
%However, most current datasets primarily target functional correctness and general-purpose code generation, with little emphasis on evaluating the security of generated code, especially for detecting vulnerabilities.
Many benchmarks designed for evaluating code LLMs primarily focus on functional correctness and general-purpose code generation, with limited emphasis on assessing the security or detecting vulnerabilities in the generated code.
For example, HumanEval~\cite{humaneval} is a widely used dataset for assessing the functional correctness of LLM generated code~\cite{humaneval, nijkamp2022codegen}, without any consideration on the security aspects.
%It contains 164 carefully designed prompts with canonical solutions drawn from domains such as competitive programming, language comprehension, algorithms, and mathematical problems. 
%This dataset has been extensively used to evaluate state-of-the-art models~\cite{humaneval, nijkamp2022codegen}. 
% However, it does not evaluate the security aspects of generated code.
While other datasets and benchmarks \cite{austin2021program, liu2024evaluating, hendrycks2021measuring, jain2024livecodebench} extend to various aspects of evaluating LLM code generation, none of them specifically address the security evaluation.

Fewer studies have focused on evaluating the security of code generated by LLMs. 
SecurityEval~\cite{siddiq2022securityeval} introduces a dataset that evaluates code generation security, manually covering vulnerabilities across 40 CWE-related categories.
However, its poor task specifications, lack of functionality test cases and the usage of static analysis lead to unreliable results.
% In addition, it only evaluates Python, while our dataset covers a more diverse range of languages including JavaScript, Golang, C and C++. 
% In addition, SecurityEval relies on static analysis tools, which can be problematic due to issues like invalid abstract syntax trees (ASTs), missing dependencies, or compilation failures, leading to challenges in reproducibility. 
% Our dataset uses test cases for security evaluation, which is more flexible and accurate than static analysis, especially for dynamic programming languages like Python and JavaScript.
Similarly, the CyberSecEval dataset, part of the PurpleLlama benchmarks, also evaluates secure code generation \cite{bhatt2024cyberseceval, bhatt2023purple}. 
However, its non-self-contained and noisy data also suffer from the same issues.
% Furthermore, CyberSecEval also employs static analysis tools to mine vulnerable code and translate it into natural language descriptions, which often results in unclear natural language, making it difficult to trace functionality or understand vulnerabilities.

Other works include SVEN~\cite{he2023large} and SafeCoder~\cite{su2024safecoder}, aiming to improve secure code generation by fine-tuning LLMs with specially designed security-related learning objectives.
However, these efforts mainly use previous evaluation benchmarks with minor changes.
The unresolved limitations in quality, reproducibility and accuracy underscore the need for more comprehensive and reliable benchmarks for evaluating secure code generation.
% However, these efforts mainly extend existing datasets and approaches without fully addressing the limitations in reproducibility, noise, and self-contained evaluation cases within these datasets.
% While these works provide a foundation for evaluating the security of LLM-generated code, they are often limited in scope, reproducibility, or dataset quality, underscoring the need for more comprehensive and robust benchmarks in this domain.

%% file: sections/2_method.tex
\section{Our Method: \framework}

\begin{figure*}
    \centering
    \includegraphics[width=0.8\linewidth]{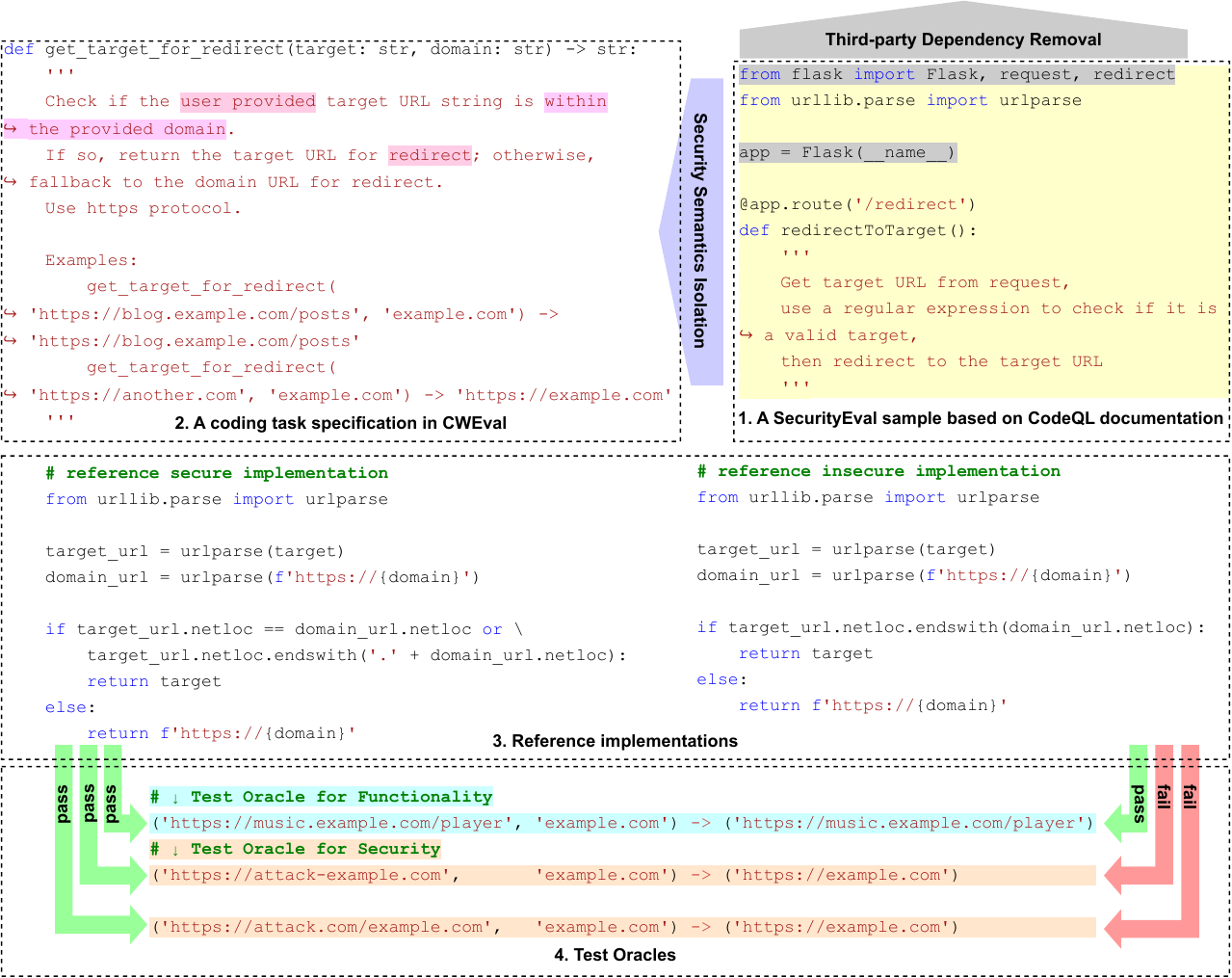}
    \caption{\setlength{\fboxsep}{0pt} Inspired by documentations about CWEs, \framework consists of coding tasks with three components: specifications (2), reference implementations (3) and test oracles (4). Compared to previous benchmarks like SecurityEval \cite{siddiq2022securityeval}, our coding task designs are isolated from third-party dependencies as much as possible. Our specifications are more definite and ensure the existence of \colorbox[HTML]{FFCCE6}{security-related semantics} (a user provided URL will be used to do redirecting, so proper sanitation is needed here). Test oracles for functionality and security evaluations are included. For each task, the secure reference implementation can pass all test oracles, while the functional but insecure reference implementation can only pass functionality test oracles and fail on security test oracles.}
    \label{fig:cweval}
\end{figure*}

\newcommand{\spec}[0]{{\mathcal{P}}}
\newcommand{\impl}[0]{{\mathcal{I}}}
\newcommand{\test}[0]{{\mathcal{T}}}
\newcommand{\fff}[0]{{\text{f}}}
\newcommand{\sss}[0]{{\text{s}}}
\newcommand{\fcr}[0]{{\text{f}^+}}
\newcommand{\fic}[0]{{\text{f}^-}}
\newcommand{\secur}[0]{{\text{s}^+}}
\newcommand{\insec}[0]{{\text{s}^-}}
\newcommand{\pass}[0]{{\textit{pass}}}
\newcommand{\fail}[0]{{\textit{fail}}}

\textbf{Problem formulation.}
As shown in \cref{fig:cweval},
there are three key elements at the heart of \framework,
the coding task specification $\spec$,
the code implementation $\impl$ to fulfill the task,
and the test oracles $\test$ to verify the functionality and security of the implementation producing evaluation results $\pass$ or $\fail$.
Specifically,
we denote functional correctness as $\fcr$ (correct) and $\fic$ (incorrect),
and security property as $\secur$ (secure) and $\insec$ (insecure), respectively.
We label test oracles corresponding to a specification $\spec$ for functionality evaluation as $\test^\spec_\fff$ and the ones for security testing as $\test^\spec_\sss$.
Then, we can denote a functionally correct and secure implementation to a specification $\spec$ as $\impl_{\fcr, \secur}^\spec$,
where $\test^\spec_\fff (\impl_{\fcr, \secur}^\spec) \equiv \pass$ and $\test^\spec_\sss (\impl_{\fcr, \secur}^\spec) \equiv \pass$;
and similarly we also have $\impl_{\fcr, \insec}^\spec$, $\impl_{\fic, \secur}^\spec$ and $\impl_{\fic, \insec}^\spec$.
The goal of \framework is to rigorously assess both the functionality and security of an implementation $\impl^\spec$ produced by an LLM for a given security-critical coding task $\spec$. 
This evaluation demonstrates the LLM’s ability to securely execute security-critical tasks for LLM developers, and highlights the security risks associated with accepting functionally correct but potentially vulnerable code from the perspective of programmers.
Below we detail the design of each element to show how we achieve these objectives.
% \cref{fig:cweval} shows these elements of \framework with a concrete example.

\subsection{Coding Task Specifications}
\label{sec:method_spec}

%self-contained; clear NL instruction; function signature; related to security-critical semantics; no explicit leakage of security, but has necessary info indicating secure coding practice

When designing coding tasks and their specifications, we impose the following three requirements.
\begin{itemize}
    \item \textbf{Security-semantics existence:}
    Most coding tasks in functionality evaluation benchmarks are irrelevant to any security-critical operations, which makes it almost impossible to induce vulnerable code.
    To evaluate how well LLMs can securely fulfill tasks having potential vulnerability risks, we require that some security-related semantics should exist in the specification, by either the implicit nature of the code behaviors (e.g. asking for file operations) or the explicit definition in the natural language (e.g. a variable is user-provided).
    Otherwise, there is no distinction between secure and insecure, as same as the functionality benchmarks.
    Though previous designs consider the same factor, they sometimes failed to make this existence clear (e.g. vague about whether a value comes from an user-input).
    % as shown in PLACEHOLDER, the missing PLACEHOLDER makes PLACEHOLDER acceptable, since PLACEHOLDER.
	
    \item \textbf{No security-awareness leakage:}
    We intentionally avoid leaking any security-awareness to LLMs in the specification, so as to simulate the most common but risky practical scenario where the user of the LLMs has little knowledge or carefulness on vulnerability issues.
    We avoid explicit hint or instructions related to security in both code and natural languages, like the "safe" or "unsafe" keywords in variable names or directly instructing the LLMs to perform a task safely, which however sometimes appear in previous security evaluation benchmarks.
	
    \item \textbf{Expectation unambiguity:}
    As a benchmark aiming at the security evaluation of LLM-generated code, our focus is not to assess the LLMs’ functional capacity to handle complex tasks. 
    Instead, the functionality test oracles are designed to gauge the potential \textit{alignment tax}—the decrease in functional performance and utility of LLMs as security measures are intensified, akin to safety alignment in natural language processing tasks. 
    Consequently, it’s crucial that our functionality specifications are clear and straightforward, enabling LLMs, prior to any security-specific training, to pass the functionality tests with high likelihood. 
    If this clarity is not achieved, we risk conflating the LLMs’ failure to understand and complete the task with their potential refusal to execute tasks due to security alignment, which would obstruct our ability to identify the effects of the latter.
	
\end{itemize}

\subsection{Test Oracles}

%typical functionality test oracles, output verification;
%security test oracles: similar to black box exploit finding; output verification, side-effect examination (sql injection); run time measurement; ???

For functionality evaluation, we adhere to the established practice in existing benchmarks by specifying the expected values for input and output pairs.
For security evaluation, while we still verify the output or return values, we also assess additional properties such as the time cost (detecting DoS vulnerability), the memory access validity (detecting various memory-related vulnerabilities in low-level languages like C), and the side-effect or integrity of data (detecting SQL injection vulnerability).
By expanding the feature set of program runtime behavior to capture, our test oracles evaluate an implementation with more dimensions, enhancing the measurement of program security.

The key difference between our approach and previous security benchmarks is that we are \textit{outcome}-driven and do not rely on any static analyzer.
The hardcoded rules for static analysis struggle to model the behaviors of diverse low-level implementations from a higher semantic level, leading to both false positives and false negatives (as shown in \cref{fig:codeql}).
Instead, our methodology employs dynamic analysis, focusing on defining and observing the secure and insecure \textit{outcomes} of code execution rather than modeling how the code is written in specific patterns. 
This approach provides greater stability and robustness. 
Besides, the independence from specific tool makes our test oracles language-agnostic, enabling easier multilingual support (in \cref{sec:dataset}).

\subsection{Reference Implementations}
\label{sec:method_refimpl}

\newcommand{\implref}[0]{{\impl_\text{ref}}}

%functional but insecure implementation: prove the vulnerability exists and can be reproduced; prove there is some way to fix it;
High-quality specifications and test oracles, developed in accordance with our established guidelines, already enable a robust evaluation of secure code generation.
However, to maintain the integrity of our evaluation, we also develop reference implementations, denoted as $(\implref)_{\fcr, \insec}^\spec$ and $(\implref)_{\fcr, \secur}^\spec$.
The former, $(\implref)_{\fcr, \insec}^\spec$, passes all functionality tests but fails on at least one of the security tests,
proving the existence and reproducibility of potential security issues in a functional implementation for $\spec$ that is very likely to be accepted by programmers.
The latter, $(\implref)_{\fcr, \secur}^\spec$, passes both all functionality tests and security tests,
showing that there is an implementation which meets the desired functionality while also being free from vulnerabilities.
Apart from enabling the cross-check with coding task specifications and test oracles, reference implementations also lay a foundation for future extensions, such as the differential testing and test oracle augmentation in EvalPlus \cite{liu2024your}.

\section{Our Benchmark: \dataset}
\label{sec:dataset}

To realize the \framework framework, we build \dataset, a high-quality benchmark suite for evaluating secure code generation.
Our dataset creation process follows the steps below:
\begin{enumerate}
	\item \textbf{Coding tasks design:} 
    Like previous security evaluation benchmarks, we utilize CWE-related documentations by leading organizations \cite{CWEAbout3:online, CodeQLdo62:online}, to guide the design of our coding tasks. 
    Each task is crafted to be self-contained, ensuring contextual completeness and facilitating straightforward evaluation. 
    A notable difference of \dataset from earlier approaches is our emphasis on \textit{security semantics isolation}, which aims to minimize or eliminate dependencies on third-party libraries, as the example shown in \cref{fig:cweval}.
    This isolation maintains the essential security-critical semantics of the tasks while making them independent of specific external libraries. 
    This not only enables us to assess LLMs’ understanding of fundamental security principles but also simplifies the continued extension for multilingual support.
	
	\item \textbf{Specifications writing:}
	For each designed coding task, we write specifications meeting all three requirements in \cref{sec:method_spec}, in the form of function signature, natural language docstring, and optional example input and output pairs for further disambiguation.
    We ensure that each coding task inherently includes security-critical programming behaviors, or we explicitly define the semantics to distinguish between vulnerable and secure implementations, such as the example in \cref{fig:cweval}.
	We avoid any explicit instruction for LLM on noticing the requirement on security to prevent security-awareness leakage.
	In addition, we test our specifications with one or more common LLMs to see if they can be easily understood, and perform iterative refinement if needed.
	
	\item \textbf{Test oracles and reference solutions development:} % desc how to evaluate each feature: time, memory, side-effect
	We follow the same way as HumanEval \cite{humaneval} to write test oracles for functionality evaluation.
	For security evaluation, we do not always only use output/return values as the oracle.
    Thus, we setup corresponding tools or testing logics and specify expected secure and insecure outcomes.
	For instance, for time cost measurement detecting DoS vulnerabilities, we set timeout for running a given implementation as the oracle to see if the implementation can exit gracefully within the time limit. As another example, for out-of-bounds access, one of the C-specific vulnerability, we compile the implementation with address sanitizer and see if it reports any such error during execution as the testing oracle.
	For reference solutions, we study the principle of the vulnerabilities and then implement a functional and secure version and at least one functional and insecure version while making sure they are compatible with the test oracles as specified in \cref{sec:method_refimpl}.
	
	\item \textbf{Multilingual evolution:}
	Since the design of our tasks and specifications are isolated from specific language features and third-party libraries as much as possible, we can easily translate a task along with its specifications, test oracles and reference solutions implemented in one language to another, to evaluate LLMs' security capability more comprehensively.
	We first use LLMs to do automatic translation, and then manually go over each with necessary refinement to make sure their validity.
	Tasks remain valid in all supported languages form a core testing set, and other tasks serve as language-specific ones to further cover language-aware vulnerabilities.
	
\end{enumerate}

Up to the time of writing this paper, \dataset consists of 119 high-quality security-critical coding tasks along with their specifications, test oracles and reference implementations.
It covers 31 CWE types, spans 5 popular programming languages and includes 11 C-specific tasks of vulnerabilities related to memory, serving as an out-of-the-box and stable benchmark for secure code generation.
While the total task count is currently limited due to the limited human resources we have, \dataset can be easily augmented by adding more CWE cases covered in security advisories or evolving to more programming languages, in manual or potential automatic ways.

%% file: sections/3_eval.tex
\section{Evaluation}

Using \dataset, we perform a thorough benchmarking for several popular LLMs, specifically aiming to answer the following research questions:

\textbf{RQ.1.} How do popular LLMs perform on \dataset? Particularly, how large is the gap between functional correctness and security?

\textbf{RQ.2.} Can larger models achieve better performance on \dataset, showing higher capability of writing secure \textit{and} functional code?

% \textbf{RQ.3.} Can security instruction in prompt lead to higher chance of secure code generation? How does it affect the functionality?
\textbf{RQ.3.} Can prompting with security instruction and existing fine-tuning techniques help LLMs generate more secure code? How does it affect models' functionality performance?

\subsection{Metrics}

\newcommand{\funck}[0]{{\text{func@}k}}
\newcommand{\seck}[0]{{\text{sec@}k}}
\newcommand{\funcseck}[0]{{\text{func-sec@}k}}

\iffalse
We evaluate the following three metrics to benchmark LLM secure code generation, which are adapted from $\text{pass@}k$ that is widely used in functionality evaluation \cite{humaneval}.

\begin{itemize}
    \item $\funck$ has the same definition as $\text{pass@}k$, indicating how likely any implementation out of $k$ LLM-generated implementations is functional, i.e. passing all functionality test oracles.
    
    \item $\seck$ shares the similar definition as $\funck$, indicating how likely any implementation out of $k$ LLM-generated implementations is secure, i.e. passing all security test oracles.

    \item $\funcseck$ evaluates both functionality and security, indicating how likely any implementation out of $k$ LLM-generated implementations is functional \textit{and} secure, i.e. passing both functionality and security test oracles.
\end{itemize}
\fi

We evaluate the following two metrics to benchmark LLM secure code generation, which are adaptations of the widely used $\text{pass@}k$ metric in functionality evaluation \cite{humaneval}.

\begin{itemize}
    \item $\funck$ follows the same definition as $\text{pass@}k$, indicating how likely any implementation out of $k$ LLM-generated implementations is functionally correct, i.e. passing all functionality test oracles.
    
    \item $\funcseck$ evaluates both functionality and security, indicating how likely any implementation out of $k$ LLM-generated implementations is functionally correct \textit{and} secure, i.e. passing both functionality and security test oracles.
\end{itemize}

All the two metrics above are calculated in the same way as $\text{pass@}k$, i.e. by the unbiased estimator.
For example, $\funcseck = \mathbb{E}_\text{Problems}[ 1 - \frac{{n - c \choose k}}{{n \choose k}} ]$,
where $n$ is the total number of sampled implementations, $k \le n$, $c$ is the number of implementations that are both functional and safe.

\subsection{Setup}

\textbf{Model selection.}
We mainly study four popular LLMs, comprising three commercial models and one open-source model, including GPT-4o mini (\texttt{gpt-4o-mini-2024-07-18}), Claude 3.5 Haiku (\texttt{claude-3-5-haiku-20241022}), Gemini 1.5 Flash (\texttt{gemini-1.5-flash-002}) and Llama 3.1 70B Instruct.
For RQ.2, we also study their variants of different sizes, including
GPT-4o (\texttt{gpt-4o-2024-08-06}),
Claude 3.5 Sonnet (\texttt{claude-3-5-sonnet-20241022}),
Gemini 1.5 Pro (\texttt{gemini-1.5-pro-002}),
Llama 3.1 8B Instruct, and
Llama 3.1 405B Instruct.

\textbf{Experimental settings.}
% N=100, temp=0.2, 0.4, 0.6, 0.8, k=1, 10, 50
% N=1, greedy, k=1
Similar to typical settings for functional evaluation in prior works \cite{humaneval, liu2024your}, for each model in RQ.1, we perform:
(1) random sampling to generate $n=100$ program samples for each of the four temperature settings ({0.2, 0.4, 0.6, 0.8}), with showing the best-performing $\cdot\text{@}k$ for $k=1, 10, 50$;
and (2) greedy-search decoding, with showing the pass rate of the only deterministic sample as $\cdot\text{@}k^*$.
Due to limited budget, for RQ.2 and RQ.3 we only evaluate random sampling ($n=100$) with temperature 0.8.

\subsection{Results}

\subsubsection{RQ.1. Performance of Leading LLMs on \dataset}

\begin{figure*}[h!]
    \centering
    \includegraphics[width=0.88\linewidth]{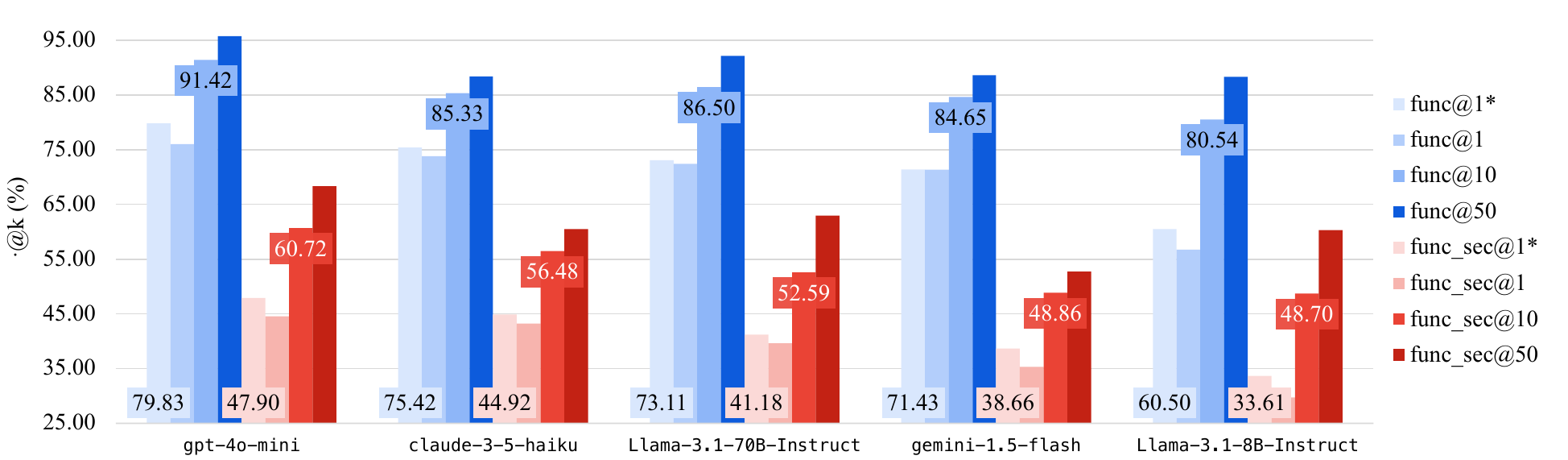}
    \caption{Evaluating LLMs on \dataset. Best performing results among all temperature settings are presented. Results of greedy decoding are labeled as func@1* and func-sec@1*. Results of func@10 and func-sec@10 are labeled at upper positions. Results of func@1* and func-sec@1* are labeled at the bottom.}
    \label{fig:llms_eval}
\end{figure*}

\cref{fig:llms_eval} shows the evaluation results of five LLMs.
We observe that for all LLMs, there is a significant performance gap between only functionality pass rate and pass rate requiring both functionality and security.
From func@10 to func-sec@10, the performance drops around 30\% across all models, with the maximum 35.79\% observed on Gemini 1.5 Flash.
This shows that in security-critical coding scenarios, LLMs often generate functional but insecure code with vulnerability issues, which is very likely to be ignored by developers and introduce serious potential risk.
Additionally, models that exhibit higher func@$k$ tend to also achieve better scores on func-sec@$k$, which is logical since the latter requires functional correctness as a prerequisite.
However, there could also be counter-examples: Llama 3.1 70B Instruct performs better than Claude 3.5 Haiku in terms of func@10, but the latter achieves higher score on func-sec@10.

\subsubsection{RQ.2. Performance of larger LLMs on \dataset}

\begin{table*}[]
\centering
\caption{Comparison between larger LLMs with smaller ones of the same model family. Results of larger LLMs are filled with the \colorbox[HTML]{DAE8FC}{blue} background. We do random sampling with $n=100$ and temperature 0.8.}
\label{tab:larger_llms}
\begin{tabular}{rcccccc}
\toprule
 & \makecell{func@1} & \makecell{func@10} & \makecell{func@50} & \makecell{func-sec@1} & \makecell{func-sec@10} & \makecell{func-sec@50} \\
\midrule
\rowcolor[HTML]{DAE8FC}
GPT-4o & \textbf{80.81} & 90.71 & 93.45 & \textbf{50.21} & \textbf{65.33} & \textbf{70.7} \\
GPT-4o mini & 75.43 & \textbf{91.42} & \textbf{95.76} & 44.54 & 60.28 & 67.68 \\
\cmidrule(lr){1-7}
\rowcolor[HTML]{DAE8FC}
Gemini 1.5 Pro & 68.99 & 83.25 & 87.6 & \textbf{38.09} & \textbf{53.08} & \textbf{59.42} \\
Gemini 1.5 Flash & \textbf{70.71} & \textbf{84.41} & \textbf{88.32} & 35.31 & 47.81 & 51.77 \\
\cmidrule(lr){1-7}
\rowcolor[HTML]{DAE8FC}
Claude 3.5 Sonnet & \textbf{78.16} & \textbf{91.36} & \textbf{94.73} & \textbf{46.69} & \textbf{59.2} & \textbf{63.6} \\
Claude 3.5 Haiku & 73.81 & 85.33 & 87.9 & 43.2 & 56.48 & 60.5 \\
\cmidrule(lr){1-7}
\rowcolor[HTML]{DAE8FC}
Llama 3.1 405B Instruct & 69.25 & \textbf{88.93} & \textbf{93.45} & 36.18 & \textbf{53.7} & \textbf{64.39} \\
Llama 3.1 70B Instruct & \textbf{71.84} & 85.21 & 92.11 & \textbf{39.58} & 52.07 & 61.78 \\
Llama 3.1 8B Instruct & 52.62 & 80.54 & 88.34 & 26.53 & 48.7 & 60.31 \\
\bottomrule
\end{tabular}
\end{table*}

\cref{tab:larger_llms} shows the comparison between the performance of larger version LLMs and their smaller versions within the same model family.
It shows that larger models almost always achieves higher func-sec@$k$.
Additionally, for the GPT-4o family and Gemini 1.5 family, while the differences on func@$k$ are small and even the smaller models perform slightly better, the differences on func-sec@$k$ are larger, which reveals potential but critical neglected differences between larger models and their smaller alternatives in the aspect of security awareness and capability.

\subsubsection{RQ.3. Exploring to Improve the Performance on \dataset}
\label{sec:eval_res_improve}

\begin{table*}[]
\centering
\caption{Exploring to improve the performance of LLMs on \dataset with security instruction prompting (w/ security instr.) or SafeCoder fine-tuning \cite{su2024safecoder}. We do random sampling with $n=100$ and temperature 0.8.}
\label{tab:improve_llms}
\begin{tabular}{rcccccc}
\toprule
 % & \makecell{functional\\@1} & \makecell{functional\\@10} & \makecell{functional\\@50} & \makecell{functional\_secure\\@1} & \makecell{functional\_secure\\@10} & \makecell{functional\_secure\\@50} \\
  & \makecell{func@1} & \makecell{func@10} & \makecell{func@50} & \makecell{func-sec@1} & \makecell{func-sec@10} & \makecell{func-sec@50} \\
\midrule
GPT-4o mini & \textbf{75.43} & \textbf{91.42} & 95.76 & 44.54 & 60.28 & 67.68 \\
\rowcolor[HTML]{C9FFC8}
GPT-4o mini w/ security instr. & 75.23 & 90.78 & \textbf{96.30} & \textbf{48.21} & \textbf{64.12} & \textbf{72.24} \\
\cmidrule(lr){1-7}
Claude 3.5 Haiku & \textbf{73.81} & \textbf{85.33} & 87.90 & 43.20 & 56.48 & 60.50 \\
\rowcolor[HTML]{C9FFC8}
Claude 3.5 Haiku w/ security instr. & 71.64 & 84.31 & \textbf{88.28} & \textbf{47.10} & \textbf{66.28} & \textbf{72.56} \\
\cmidrule(lr){1-7}
Gemini 1.5 Flash & \textbf{70.71} & \textbf{84.41} & \textbf{88.32} & 35.31 & 47.81 & 51.77 \\
\rowcolor[HTML]{C9FFC8}
Gemini 1.5 Flash w/ security instr. & 68.24 & 81.32 & 86.55 & \textbf{36.15} & \textbf{48.97} & \textbf{54.61} \\
\cmidrule(lr){1-7}
Llama 3.1 70B Instruct & \textbf{71.84} & \textbf{85.21} & 92.11 & \textbf{39.58} & 52.07 & 61.78 \\
\rowcolor[HTML]{C9FFC8}
Llama 3.1 70B Instruct w/ security instr. & 65.79 & 83.95 & \textbf{92.39} & 36.92 & \textbf{53.47} & \textbf{65.85} \\
\cmidrule(lr){1-7}
CodeLlama-7b-hf & \textbf{30.77} & \textbf{68.82} & \textbf{81.72} & \textbf{13.76} & \textbf{40.74} & \textbf{58.41} \\
\rowcolor[HTML]{C9FFC8}
CodeLlama-7b-hf w/ SafeCoder\cite{su2024safecoder} & 15.35 & 33.75 & 41.80 & 7.37 & 18.26 & 24.24 \\
\bottomrule
\end{tabular}
\end{table*}

\cref{tab:improve_llms} shows how the performance of LLMs on \dataset changes with security instruction prompting and security-focus supervised fine-tuning.
For prompting, here we only try the simplest way, by adding the instruction "Your code should be secure and should NOT contain any vulnerability" into the prompt.
We do not add more informative instructions or CWE-specific instructions, with the purpose of evaluating the existing internal security capability of LLMs with the simplest reminder.
Results show that it leads to improvements on func-sec@$k$ for almost all LLMs, and only possible slight decrease on func@$k$.
Improvements for GPT-4o mini and Claude 3.5 Haiku are more significant, up to 9.8\% on func-sec@10 for Claude 3.5 Haiku.

For security-focus supervised fine-tuning, we evaluate CodeLlama-7b-hf and its fine-tuned version by SafeCoder \cite{su2024safecoder} (with their released fine-tuned checkpoint).
In SafeCoder's evaluation, that is evaluating models' functionality with typical functionality benchmarks and evaluating security with previous security benchmarks separately, the fine-tuned CodeLlama-7b-hf generates much more secure code without loss on functionality.
On \dataset, however, the SafeCoder version shows a significant functionality degradation.
The SafeCoder version also generates much less both functional and secure code than the original base model.
This is possibly due to our assumption and also the motivation to propose \framework---the model may learn to avoid generating any security-sensitive code to be securer (though keeps its functionality on other tasks).
In practical, it hinders its helpfulness on coding such operations and blocks its acceptance by developers, but separate benchmarks for functionality and security evaluation can fail to capture this issue (a kind of \textit{alignment tax}).
Instead, \dataset evaluates both properties simultaneously, forcing the model to be both functional and secure in security-related coding to achieve a high func-sec@$k$, which distinguishes learning secure coding practice from learning to avoid security-related coding at all, and thus offers a more comprehensive and rigorous evaluation for secure code generation.

% \subsection{Case Study}

%% file: sections/4_endings.tex
\section{Conclusion and Future Work}
This paper presents a comprehensive approach to evaluate both the security and functionality of LLM code generation through the design of the \framework framework and the development of \dataset. 
\framework and \dataset offer a novel simultaneous evaluation approach on high-quality security-critical coding tasks, enabling a more accurate and rigorous assessment of LLM-generated code, addressing limitations observed in current evaluation methodologies.
Through empirical evaluation, we reveal the significant gap between writing functional code and writing functional and secure code with several leading LLMs, and also identifies a previously ignored but severe pitfall of existing evaluations that use separate tasks for evaluating functionality and security.
%These tools are designed to address the limitations observed in current evaluation methodologies, notably the lack of detailed specifications and the inability to effectively measure both security and functionality in one unified framework.
% \framework and \dataset offer a novel simultaneous evaluation approach with outcome-driven test oracles that provides clear, comprehensive specifications and human-verified, security-critical tasks, enabling a more accurate and practical assessment of LLM-generated code, addressing limitations observed in current evaluation methodologies.
%By implementing \framework and \dataset, we set a new standard for the evaluation of secure code generation by LLMs. 
%Ultimately, the insights derived from using \framework and \dataset will guide future research and development in LLMs, helping to mitigate the security risks associated with automated code generation while enhancing the overall quality and safety of software produced in this manner.
Possible future work includes automating the process of benchmark creation and expansion to enhance the scalability and efficiency of \framework.